\documentclass[twocolumn,english,aps,superscriptaddress,showpacs,pre,floatfix,nofootinbib,longbibliography]{revtex4-1}

\usepackage[latin9]{inputenc}
\usepackage{color}
\usepackage{babel}
\usepackage{amsmath}
\usepackage{amssymb}
\usepackage{graphicx}
\usepackage{wasysym}
\usepackage{esint}
\usepackage[unicode=true,pdfusetitle,
 bookmarks=true,bookmarksnumbered=false,bookmarksopen=false,
 breaklinks=true,pdfborder={0 0 0},pdfborderstyle={},backref=false,colorlinks=true]
 {hyperref}
\hypersetup{
 linkcolor=blue,citecolor=blue,urlcolor=blue}

\makeatletter
\usepackage{babel}
\usepackage{amsmath}

\makeatother

\begin{document}
\title{Dynamic quantum phase transitions in the two-leg Creutz ladder with
long-range hopping }
\author{J. A. da Silva }
\affiliation{Universidade Federal de Uberlândia, Instituto de Física, C.~P.~593,
38400-902 Uberlândia, MG, Brazil}
\author{J. C. Xavier}
\affiliation{Universidade Federal de Uberlândia, Instituto de Física, C.~P.~593,
38400-902 Uberlândia, MG, Brazil}
\begin{abstract}
In this work, we investigate quantum quenches in the two-leg Creutz
model with long-range hopping, where the hopping amplitudes decay
with distance as a power law characterized by an exponent $\nu$ and
have a finite range $D$. We first obtain the exact solution of a
generic two-band model in momentum space. This allows us to compute
the Loschmidt amplitude and, consequently, the dynamical free energy
$f(\texttt{t})$ of the two-band model. We also show how to determine
the Yang-Lee-Fisher (YLF) zeros by solving a nonlinear equation. We
demonstrate that the two-leg Creutz model in momentum space is a special
case of the generic two-band model. Using these results, we identify
the non-analyticities in the dynamical free energy $f(\texttt{t})$
at critical times $\texttt{t}_{c}$. We find that the number of nontrivial
critical times $N_{s}$ depends on both $\nu$ and $D$. In particular,
we show that for small $\nu$ and large $D$ the critical times become
increasingly dense, leading, in the appropriate regime, to non-analyticities
at an increasingly dense set of times---similar to what was observed
by Xavier and Hoyos {[}Phys. Rev. B, \textbf{108}, 214303 (2023){]}
in the Su-Schrieffer-Heeger (SSH) model with long-range hopping terms. 
\end{abstract}
\date{\today}
\maketitle

\section{Introduction}

It is common to characterize the phase of a system by one of its physical
properties, which may or may not be structural. Many systems undergo
a phase transition when a parameter $p$ (e.g., temperature) exceeds
or falls below a critical value $p_{c}$. At $p_{c}$, the physical
properties of the system typically exhibit some type of singularity
or anomaly. For this reason, understanding phase transitions (PTs)
has long been a subject of great interest. Equilibrium phase transitions,
for example, have been extensively studied over the past century and
are now relatively well understood \citep{booksachdev,bookStanley,RevPT1997}.

In equilibrium PTs, the different phases are characterized by an order
parameter (e.g., the density), which takes distinct values in each
phase. The phase boundaries are marked by non-analyticities in the
thermodynamic functions, which are encoded in the zeros of the partition
function ${\cal Z}$, the so-called Yang-Lee-Fisher (YLF) zeros~\citep{YLzerosI,YLzerosII,bookFisherYLzeros}.
In general, the zeros of ${\cal Z}(q)=\text{Tr}\left(e^{-qH}\right)$
occur for complex values of the inverse temperature $q=\beta+i\alpha$,
where $\beta=\frac{1}{k_{B}T}$ and $\alpha\neq0$. In the thermodynamic
limit, these zeros may reach the real inverse-temperature axis, giving
rise to non-analyticities in the Helmholtz free energy $F=-k_{B}T\ln\left[{\cal Z}(\beta)\right]$.
Therefore, one way to identify an equilibrium PT is to determine the
YLF zeros and verify whether they reach the real axis. These zeros
have been experimentally observed in recent years ~\citep{PRLYLFzeros2012,PRLYLFzeros2015}.

Dynamical phase transitions in nonequilibrium quantum systems have
also been investigated. To the best of our knowledge, some of the
earliest uses of the term \emph{dynamical phase transition} in quantum
many-body systems can be found in Refs. \citealp{PhysRevA.72.052319,PhysRevLett.97.200601,QDPT-Gonzalo},
where dynamical critical behavior was characterized through the evolution
of physical quantities following a change in the system parameters.
A distinct notion of a dynamical quantum phase transition (DQPT),
based on an analogy with equilibrium PTs, was introduced in the seminal
work of Heyl, Polkovnikov, and Kehrein ~\citep{HeylPRLseminalDynamic}.
In this case, non-analytic behavior appears at critical times $\texttt{t}_{c}$
in the dynamical free energy 
\begin{equation}
f(\texttt{t})=-\frac{1}{N}\ln\left(\left|Z(\texttt{t})\right|^{2}\right),
\end{equation}
where $Z(\texttt{t})=\left\langle \psi_{0}\left|e^{-iH(\delta)\texttt{t}}\right|\psi_{0}\right\rangle $
is the Loschmidt amplitude, $\left|\psi_{0}\right\rangle $ is the
ground state of the Hamiltonian $H\left(\delta_{0}\right)$, and $N$
is the number of degrees of freedom. Here, a sudden quantum quench
is considered: a parameter $\delta$ of the Hamiltonian changes instantaneously
from $\delta_{0}$ to $\delta$ at time $\texttt{t}=0$. It is worth
mentioning that the Loschmidt amplitude is closely related to the
equilibrium boundary partition function $\mathcal{Z}^{b}(\beta)=\left\langle \psi^{b}\left|e^{-\beta H}\right|\psi^{b}\right\rangle $,
associated with the Hamiltonian $H$ and boundary conditions specified
by the boundary state $\left|\psi^{b}\right\rangle $, separated by
an imaginary-time interval $\beta$~\citep{Cardyboundary,cardyboundary2,LECLAIboundary}.

Later, a distinction between two types of dynamical phase transitions
was introduced in the literature \citep{DynamicLongRangePRB2017}.
Dynamical transitions characterized by the behavior of an order parameter
in a (quasi-)steady state were termed type-I dynamical phase transitions
(DPT-I), whereas transitions characterized by non-analyticities of
the Loschmidt return rate were termed type-II dynamical phase transitions
(DPT-II). In this terminology, the early dynamical transitions discussed
above are conceptually closer to DPT-I, whereas the notion introduced
by Heyl, Polkovnikov, and Kehrein corresponds to DPT-II.

In the following, we use the term dynamical quantum phase transition
(DQPT) to refer specifically to DPT-II, in which the dynamical free
energy exhibits non-analytic behavior at a critical time $\texttt{t}_{c}$
\citep{HeylPRLseminalDynamic,dynampt}. DQPTs have been extensively
studied in a wide variety of systems \citep{HeylPRLseminalDynamic,NetoRafaelXavierPRBL2022,KarraschPhysRevB.87.195104,DynamicalMendoza-Arenas,dynamicDelgadoPhysRevB.99.054302,DynamicVajnaPRB2014,DynamSirkerPRB2014,FirtOderDynanimcPRL,PRBVajnaDynamic,DynamicalSacramento,DynamicalVanhala,KitaevLongRangQDPT2017,KitaevLongRangeHalimeh,KitaevLongHalimedsupercon,PRLHeylLongRangDynamic,DynamicLongRangePRB2017,DQPTlograngisingPhysRevB.96.104436,DQPTlongrangePhysRevE.96.062118,DynamicalLongRangeSyed,DynamicalLongKitaev,Xavier-DQPT-LR,DynamicalCao,dynaphotonic},
and it is now well established that there is no one-to-one correspondence
between DQPTs and equilibrium phase transitions~\citep{DynamSirkerPRB2014,DynamicVajnaPRB2014,FirtOderDynanimcPRL,PRBVajnaDynamic,DynamicLongRangePRB2017,PRLHeylLongRangDynamic,SciRepJafari}.

DQPTs have also been observed experimentally. For example, trapped-ion
quantum simulators have been used to simulate the transverse-field
Ising chain with long-range interactions ~\citep{heyl-trappexp,DQPTexpPhysRevApplied.11.044080,DQPTexpnatur2018,dynaphotonic}.
From a theoretical perspective, the effects of long-range interactions
and hopping on DQPTs have been investigated in several systems \citep{KitaevLongRangQDPT2017,KitaevLongRangeHalimeh,KitaevLongHalimedsupercon,DynamicalMendoza-Arenas,DynamicalSacramento,DQPTlongrangePhysRevE.96.062118,DynamicLongRangePRB2017,Xavier-DQPT-LR,PRLHeylLongRangDynamic,DQPTlograngisingPhysRevB.96.104436,DynamicalLongRangeSyed,DynamicalLongKitaev,Dynamicalexponentalrange,DynamicVajnaPRB2014,Bojan-LMGmodel,Sacramento2}.
In particular, Ref. \citealp{Xavier-DQPT-LR} showed that, in the
Su-Schrieffer-Heeger (SSH) chain~\citep{SShmodel} with long-range
hopping, the number of times at which the Yang-Lee-Fisher (YLF) zeros
reach the real-time axis increases with the hopping range $D$. Moreover,
these critical times become nearly uniformly distributed within a
short-time interval, leading to non-analyticities at almost all times.
A similar phenomenon, although originating from a different mechanism,
was also reported in disordered systems exhibiting dynamical Griffiths
singularities \citep{NetoRafaelXavierPRBL2022}.

In the present work, we investigate the effects of long-range hopping
on DQPTs in the Creutz model \citep{PRL-CreutzModel} and examine
whether the emergence of non-analyticities at almost all times is
a generic consequence of long-range hopping or a feature specific
to the SSH model. Like the SSH chain, the Creutz model is an exactly
solvable free-fermion model defined on a two-leg ladder subject to
a uniform magnetic field \citep{PRL-CreutzModel}. This exact solvability
allows us to derive analytical expressions for the critical times.

A theoretical study of DQPTs in the nearest-neighbor Creutz model
was previously carried out in Ref. \citep{dynamicDelgadoPhysRevB.99.054302}.
However, the effects of long-range hopping were not addressed. Furthermore,
the Creutz model has recently been realized experimentally in a one-dimensional
optical lattice \citep{Creutzopticallattice}. This experimental progress
makes the investigation of DQPTs in the long-range Creutz model particularly
timely. We expect that our results will contribute to a better understanding
of how long-range hopping influences dynamical quantum phase transitions
and may provide guidance for future experimental studies.

The paper is organized as follows. In Sec.~\ref{sec:twobandModel},
we present a generic two-band model in momentum space and its exact
diagonalization. Analytical expressions for the Loschmidt amplitude,
dynamical free energy, and the YLF zeros are also presented in this
section. In Sec.~\ref{sec:Creutz}, we show how a uniform magnetic
field affects the long-range hopping amplitudes in the Creutz model
and then investigate the DQPTs of this model. Finally, we present
our concluding remarks in Sec.~\ref{sec:CONCLUSION}.

\section{DQPTs of a generic two-band model \label{sec:twobandModel}}

\subsection{A Generic Two-Band Model\label{subsec:2band} }

Let us consider the following generic spinless two-band model in momentum
space 
\begin{eqnarray}
H & = & \sum_{k}\left(\begin{array}{cc}
A_{k}^{\dagger} & B_{k}^{\dagger}\end{array}\right)\begin{pmatrix}M_{11}(k,\theta) & M_{12}(k,\theta)\\
M_{21}(k,\theta) & M_{22}(k,\theta)
\end{pmatrix}\left(\begin{array}{c}
A_{k}^{\phantom{\dagger}}\\
B_{k}^{\phantom{\dagger}}
\end{array}\right),\nonumber \\
 & = & \sum_{k}\varGamma_{k}^{\dagger}M(k,\theta)\varGamma_{k},\label{eq:H2band}
\end{eqnarray}
where $A_{k}$ and $B_{k}$ are the fermionic annihilation operators,
$\varGamma_{k}^{\dagger}=\left(\begin{array}{cc}
A_{k}^{\dagger} & B_{k}^{\dagger}\end{array}\right)$, and $\theta$ is a parameter of the Hamiltonian. We assume that
the matrix $M$, whose elements are $M_{i,j}$, is Hermitian, and
that the momenta $k$ are given by $k=k_{n}=\frac{2\pi}{L}n$, with
$n=0,1,2,\dots,L-1$.

The above Hamiltonian can be easily diagonalized, yielding 
\begin{equation}
H=\sum_{k}\left(\omega_{k}^{-}\alpha_{k}^{\dagger}\alpha_{k}+\omega_{k}^{+}\beta_{k}^{\dagger}\beta_{k}\right),
\end{equation}
where the eigenenergies are

\begin{equation}
\omega_{k}^{\pm}(\theta)=\frac{1}{2}\left(M_{11}+M_{22}\pm\sqrt{(M_{11}-M_{22})^{2}+4M_{21}M_{12}}\right),\label{eq:dispertion2band}
\end{equation}
and the eigenoperators associated with the positive and negative branches
of the dispersion relation $\omega_{k}^{\pm}(\theta)$ are

\begin{equation}
\begin{cases}
\alpha_{k}^{\dagger}(\theta)= & \cos\biggl(\frac{\gamma_{k,\theta}}{2}\biggr)A_{k}^{\dagger}-\sin\biggl(\frac{\gamma_{k,\theta}}{2}\biggr)B_{k}^{\dagger}\thinspace,\\
\beta_{k}^{\dagger}(\theta)= & \sin\biggl(\frac{\gamma_{k,\theta}}{2}\biggr)A_{k}^{\dagger}+\cos\biggl(\frac{\gamma_{k,\theta}}{2}\biggr)B_{k}^{\dagger}\thinspace,
\end{cases}\label{eq:OperAlpha}
\end{equation}
with 
\begin{equation}
\tan(\gamma_{k,\theta})=\frac{-2M_{12}(k,\theta)}{M_{11}(k,\theta)-M_{22}(k,\theta)}.\label{eq:phasegamma}
\end{equation}
The ground state of $H(\theta)$, for a system at half-filling, is
\begin{equation}
\left|\psi_{0}(\theta)\right\rangle =\prod_{k}\alpha_{k}^{\dagger}\left|0\right\rangle ,
\end{equation}
where the product runs over all values of $k$.

\subsection{The Loschmidt Amplitude, Dynamical Free Energy, and the YLF Zeros}

We can now evaluate the Loschmidt amplitude, since we have diagonalized
the generic two-band model and obtained its eigenenergies and eigenoperators.
Our quench protocol is as follows: the system is initialized in the
ground state $\left|\psi_{0}(\theta)\right\rangle =\prod_{k}\alpha_{k}^{\dagger}\left|0\right\rangle $
of $H(\theta)\equiv H$, and time-evolved according to $\tilde{H}\equiv H(\tilde{\theta})$.

Let us first evaluate the Loschmidt amplitude 
\begin{eqnarray}
Z(\texttt{t}) & = & \left\langle \psi_{0}(\theta)\left|e^{-iH(\tilde{\theta})\texttt{t}}\right|\psi_{0}(\theta)\right\rangle ,\nonumber \\
 & = & \left\langle 0\left|\prod_{q}\alpha_{k}e^{-iH(\tilde{\theta})\texttt{t}}\prod_{k}\alpha_{k}^{\dagger}\right|0\right\rangle ,\label{eq:LE}
\end{eqnarray}
following the same procedure as in Ref.~\citep{NetoRafaelXavierPRBL2022}.
We first need to relate the pre-quench eigenoperator $\alpha_{k}^{\dagger}\equiv\alpha_{k}^{\dagger}(\theta)$
to the post-quench eigenoperators $\tilde{\alpha}_{k}^{\dagger}\equiv\alpha_{k}^{\dagger}(\tilde{\theta})$
and $\tilde{\beta}_{k}^{\dagger}\equiv\beta_{k}^{\dagger}(\tilde{\theta})$.
By using Eq. (\ref{eq:OperAlpha}), we obtain 
\begin{equation}
\alpha_{k}^{\dagger}=\left[\cos\biggl(\eta_{k}\biggr)\tilde{\alpha}_{k}^{\dagger}-\sin\biggl(\eta_{k}\biggr)\tilde{\beta}_{k}^{\dagger}\right],
\end{equation}
where $\eta_{k}\equiv\frac{\gamma_{k,\theta}-\gamma_{k,\tilde{\theta}}}{2}$.
Now, using the fact that 
\begin{equation}
\begin{cases}
e^{-i\tilde{H}\texttt{t}}\tilde{\alpha}_{k}^{\dagger}= & e^{-i\tilde{\omega}_{k}^{-}\texttt{t}}\tilde{\alpha}_{k}^{\dagger}e^{-i\tilde{H}\texttt{t}}\thinspace,\\
e^{-i\tilde{H}\texttt{t}}\tilde{\beta}_{k}^{\dagger}= & e^{-i\tilde{\omega}_{k}^{+}\texttt{t}}\tilde{\beta}_{k}^{\dagger}e^{-i\tilde{H}\texttt{t}}\thinspace,
\end{cases}
\end{equation}
where $\tilde{\omega}_{k}^{\pm}\equiv\omega_{k}^{\pm}(\tilde{\theta})$
are the eigenenergies of $\tilde{H}$, it is straightforward to show
that 
\begin{equation}
e^{-i\tilde{H}\texttt{t}}\left|\psi_{0}\right\rangle =\prod_{k}\left[\cos\bigl(\eta_{k}\bigr)e^{-i\tilde{\omega}_{k}^{-}\texttt{t}}\tilde{\alpha}_{k}^{\dagger}-\sin\bigl(\eta_{k}\bigr)e^{-i\tilde{\omega}_{k}^{+}\texttt{t}}\tilde{\beta}_{k}^{\dagger}\right]\left|0\right\rangle .\label{eq:eHt}
\end{equation}
Finally, inserting Eq. (\ref{eq:eHt}) into Eq. (\ref{eq:LE}), we
obtain the Loschmidt amplitude for the generic Hamiltonian (\ref{eq:H2band}),
given by 
\begin{equation}
Z(\texttt{t})=\prod_{k}\left[\cos^{2}\bigl(\eta_{k}\bigr)e^{-i\tilde{\omega}_{k}^{-}\texttt{t}}+\sin^{2}\bigl(\eta_{k}\bigr)e^{-i\tilde{\omega}_{k}^{+}\texttt{t}}\right].\label{eq:LEexact}
\end{equation}

Having calculated the Loschmidt amplitude, it is straightforward to
obtain the dynamical free energy and the YLF zeros. The dynamical
free energy $f(\texttt{t})$ is

\begin{equation}
f(\texttt{t})=-\frac{1}{L}\sum_{k}\ln\left[1-\sin^{2}\bigl(2\eta_{k}\bigr)\sin^{2}\biggl(\frac{\Delta\tilde{E}_{k}}{2}\texttt{t}\biggr)\right],\label{eq:DynEner-2}
\end{equation}
where $\Delta\tilde{E}_{k}\equiv\tilde{\omega}_{k}^{+}-\tilde{\omega}_{k}^{-}$.
In the thermodynamic limit, the sum can be replaced by an integral

\begin{equation}
f(\texttt{t})=-\intop_{0}^{2\pi}\frac{dk}{2\pi}\ln\left[1-\sin^{2}\bigl(2\eta_{k}\bigr)\sin^{2}\biggl(\frac{\Delta\tilde{E}_{k}}{2}\texttt{t}\biggr)\right].\label{eq:DynEner-1-1}
\end{equation}
The non-analyticities of $f(\texttt{t})$ are determined by the YLF
zeros. Let $z=\texttt{t}+i\tau$. We then need to find the values
of $z$ such that 
\begin{equation}
Z(z)=\prod_{k}e^{-z\tilde{\omega}_{k}^{-}}\left[\cos^{2}\bigl(\eta_{k}\bigr)+\sin^{2}\bigl(\eta_{k}\bigr)e^{-zi\Delta\tilde{E_{k}}}\right]=0.
\end{equation}
Therefore, the YLF zeros are 
\begin{equation}
z_{m}(k_{n})=\frac{1}{\Delta\tilde{E}_{k}}\left[\pi\left(2m+1\right)+i\ln\left(\tan^{2}(\eta_{k_{n}})\right)\right],\label{eq:YLZ}
\end{equation}
where $m\in\mathbb{N}_{0}$ labels the $m$-th accumulation line of
YLF zeros, and $n=0,1,\dots,L-1$ labels the $n$-th momentum value
$k_{n}$. This expression is equivalent to the well-known form of
the Fisher zeros written in terms of the overlap between the pre-
and post-quench Bloch vectors \citep{HeylPRLseminalDynamic,DynamicVajnaPRB2014,PRBVajnaDynamic}.
In fact, for a two-band Hamiltonian $M(k,\theta)=d_{0}(k,\theta)I+\mathbf{d}(k,\theta)\cdot\boldsymbol{\sigma}$,
with real off-diagonal matrix elements, one has $\hat{\mathbf{d}}_{k}^{\,i}\cdot\hat{\mathbf{d}}_{k}^{\,f}=\cos(2\eta_{k})$.
Using $\operatorname{arctanh}[\cos(2\eta_{k})]=-\ln|\tan\eta_{k}|$,
Eq. (\ref{eq:YLZ}) can therefore be directly mapped onto the conventional
Bloch-vector representation \citep{PRBVajnaDynamic}. 

As we can see from Eq. (\ref{eq:YLZ}), there are two fundamental
quantities underlying the YLF zeros: the eigenenergies $\tilde{\omega}_{k}^{\pm}$
and the phases $\gamma_{k,\theta}$ and $\gamma_{k,\tilde{\theta}}$,
which are associated with the initial and final parameters of the
Hamiltonian, respectively. The non-analyticities of $f(\texttt{t})$
occur only when the YLF zeros reach the real-time axis, i.e., $z_{m}(k_{n})=\texttt{t}+0i$.

From Eq. (\ref{eq:YLZ}), we see that this happens only if there exists
a momentum $k_{n}^{\star}$ such that 
\begin{equation}
\tan^{2}(\eta_{k_{n}^{\star}})=1.
\end{equation}
Equivalently, in the Bloch-vector representation, this condition corresponds
to $\hat{\mathbf{d}}_{k^{\star}}^{\,i}\cdot\hat{\mathbf{d}}_{k^{\star}}^{\,f}=0$
\citep{DynamicVajnaPRB2014,PRBVajnaDynamic}. This occurs if $\eta_{k_{n}^{\star}}=\frac{\gamma_{k_{n}^{\star},\theta_{1}}-\gamma_{k_{n}^{\star},\theta_{2}}}{2}=\pm\frac{\pi}{4}$.
In terms of the elements of the matrix $M$, the above constraint
yields the following equation

\begin{widetext}
\begin{equation}
4M_{12}(k_{n}^{\star},\theta)M_{12}(k_{n}^{\star},\tilde{\theta})=-\biggl(M_{11}(k_{n}^{\star},\theta)-M_{22}(k_{n}^{\star},\theta)\biggr)\biggl(M_{11}(k_{n}^{\star},\tilde{\theta})-M_{22}(k_{n}^{\star},\tilde{\theta})\biggr).\label{eq:YLZ-tau0}
\end{equation}
It is interesting to note that in the case of two-band Hamiltonians
for which $M_{12}$ is independent of $\theta$, we have $M_{12}(k_{n}^{\star},\theta)M_{12}(k_{n}^{\star},\tilde{\theta})\ge0$.
Therefore, in this case, the DQPTs occur only if 
\begin{equation}
\biggl(M_{11}(k_{n}^{\star},\theta)-M_{22}(k_{n}^{\star},\theta)\biggr)\biggl(M_{11}(k_{n}^{\star},\tilde{\theta})-M_{22}(k_{n}^{\star},\tilde{\theta})\biggr)<0.
\end{equation}

\end{widetext}Assuming that solutions to Eq. (\ref{eq:YLZ-tau0})
exist, we find that the non-analyticities of the dynamical free energy
$f(\texttt{t})$ occur at the following critical times 
\begin{equation}
\texttt{t}_{k_{n}^{\star},m}=\frac{2\pi}{\Delta\tilde{E}_{k_{n}^{\star}}}\Bigl(m+\frac{1}{2}\Bigr).\label{eq:tc}
\end{equation}
It is worth noting that the present formulation, expressed directly
in terms of eigenstates, energy eigenvalues, and their overlaps, can
be naturally extended to multiband systems, for which the simple geometric
Bloch-vector representation of two-band systems becomes considerably
less straightforward. In the next section, we use these results to
investigate the DQPTs of a particular model, namely, the Creutz model.

\section{DQPTs of the two-leg Creutz ladder\label{sec:Creutz}}

\subsection{The Creutz Model with Long-Range Hopping}

\begin{figure}
\begin{centering}
\includegraphics[scale=0.4]{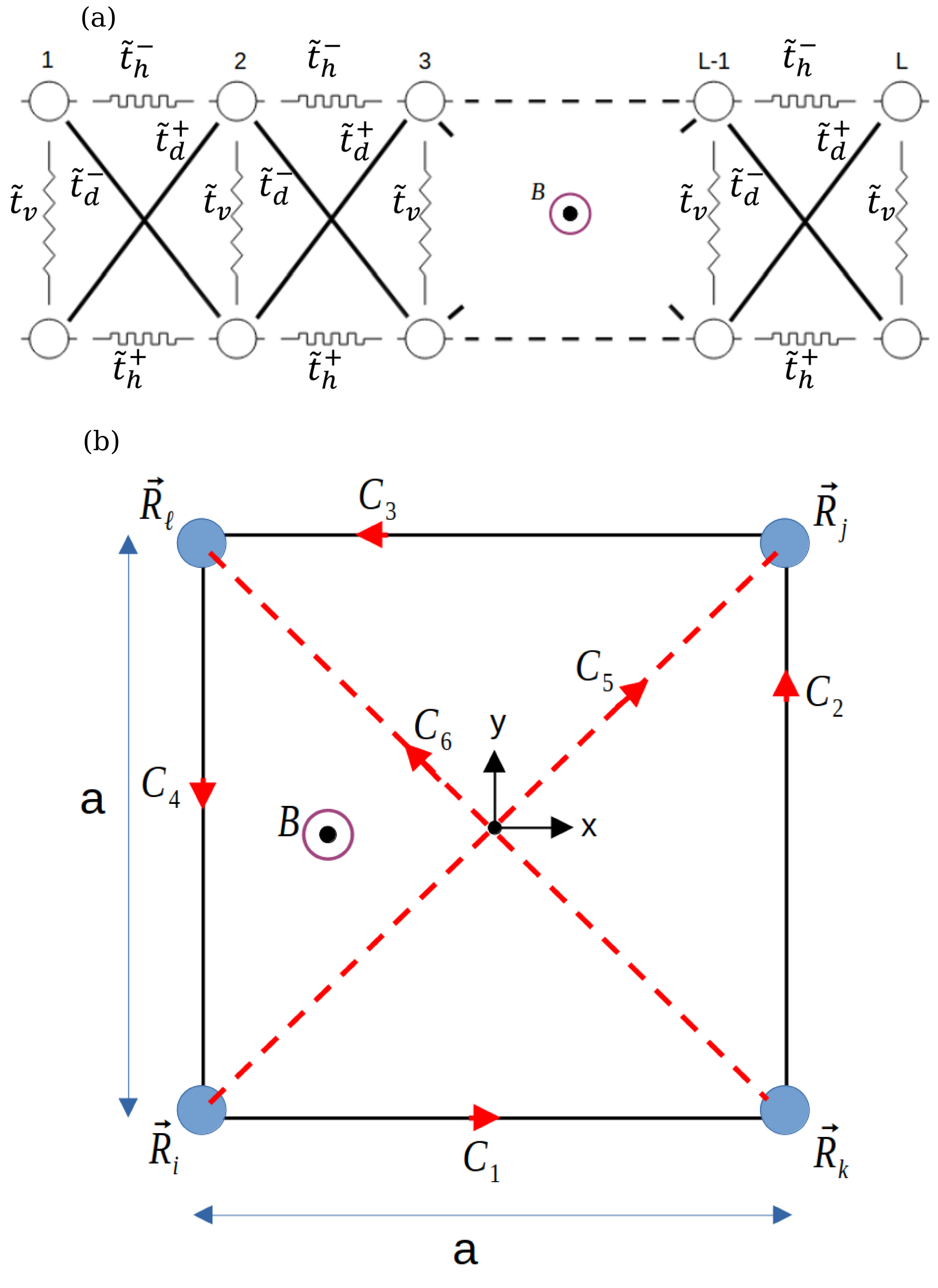}
\par\end{centering}
\caption{(a) Representation of the two-leg Creutz ladder with length $L$ and
$D=1$. The hopping amplitudes in the presence of a uniform perpendicular
magnetic field are given by $\tilde{t}_{x,1}\equiv\tilde{t}_{x}^{\pm}=e^{\pm i\theta}t_{x}^{\pm}$,
$x=h,d$, and $\tilde{t}_{v}=t_{v}$. Here $\theta=\frac{q}{\hbar}\frac{\Phi}{2},$
where $\Phi$ is the magnetic flux through a single plaquette. The
parameters $t_{x}^{\pm},$$x=h,d$, and $t_{v}$ denote the hopping
amplitudes in the absence of a magnetic field (see text). (b) Paths
used to evaluate the hopping amplitudes (see text). The lattice spacing
is set to $a=1$. \label{fig:2leg}}
\end{figure}

Consider, initially, a two-leg free-fermion ladder with long-range
hopping, in the \emph{absence of a magnetic field}, under periodic
boundary conditions, described by the Hamiltonian:

\begin{gather}
H_{0}=-\sum_{n=1}^{L}\sum_{\ell=1}^{D}\Biggl[\Bigl(t_{h,\ell}^{+}a_{n}^{\dagger}a_{n+\ell}+h.c.\Bigr)+\Bigl(t_{h,\ell}^{-}b_{n}^{\dagger}b_{n+\ell}+h.c.\Bigr)\nonumber \\
+\Bigl(t_{d,\ell}^{+}a_{n}^{\dagger}b_{n+\ell}+h.c.\Bigr)+\Bigl(t_{d,\ell}^{-}b_{n}^{\dagger}a_{n+\ell}+h.c.\Bigr)\label{eq:HB0}\\
+\Bigl(t_{v}a_{n}^{\dagger}b_{n}+h.c.\Bigr)\Biggr]\textrm{,}\nonumber 
\end{gather}
where $D$ is the hopping range, and we assume that the hopping amplitudes
decay as $t_{h,\ell}^{\pm}=\frac{t_{h}}{\ell^{\nu}}$ and $t_{d,\ell}^{\pm}=\frac{t_{d}}{\ell^{\nu}}$,
where $\nu$ is the decay exponent (see, for instance, Fig. \ref{fig:2leg}(a)
for the case $D=1$). The subscripts $h$, $v$, and $d$ refer to
the horizontal, vertical, and diagonal hopping amplitudes, respectively.
The operators $a_{n}$ and $b_{n}$ are the fermionic annihilation
operators at the $n$-th site of the lower and upper legs of the ladder
of length $L$, respectively. It is worth noting that hopping amplitudes
decaying as $t_{i,i+\ell}\sim1/\ell^{\nu}$ have also been considered
in the context of the Su-Schrieffer-Heeger (SSH) chain~\citep{SShmodel}.
For certain parameter choices, this decay has been used to study symmetry-resolved
entanglement entropy~\citep{Ares_Calabrese2022LongRange,LongRangeJones}
as well as DQPTs \citep{Xavier-DQPT-LR}.

The effect of a uniform magnetic field on Bloch electrons is reasonably
well understood (see, for instance, Refs. \onlinecite{Peierls-Subst,Luttinger-MagField,HofstadterButterfly,KohnMagField}).
However, for the sake of completeness, we briefly discuss the main
consequence of a perpendicular uniform magnetic field on the Hamiltonian
(\ref{eq:HB0}). In the case of an effective one-band model, the hopping
amplitudes are given by \citep{Luttinger-MagField}

\begin{widetext} 

\begin{equation}
\tilde{t}_{i,j}=\int d^{3}r\exp\biggl(\frac{iq}{\hbar}\varint_{\overrightarrow{R_{i}}}^{\overrightarrow{R_{j}}}\overrightarrow{A}\cdot d\overrightarrow{s}\biggr)W_{i}^{*}(\vec{r})\left(\frac{1}{2m_{e}}\left(\vec{p}-\frac{q}{c}\vec{A}\right)^{2}+U\right)W_{j}(\vec{r}),
\end{equation}

\end{widetext} where $W_{i}(\vec{r})$ is the Wannier function centered
at the site $\vec{R}_{i}$, $\vec{A}$ is the vector potential, and
$U$ is the periodic potential. The exact evaluation of this amplitude
is nontrivial. However, when the Wannier functions are well localized
around their respective sites and the magnetic field is sufficiently
strong, the hopping amplitudes can be approximated using the Peierls
substitution \citep{Peierls-Subst,Luttinger-MagField}
\begin{equation}
\tilde{t}_{i,j}(\vec{A})=t_{i,j}\exp\biggl(\frac{iq}{\hbar}\varint_{\overrightarrow{R_{i}}}^{\overrightarrow{R_{j}}}\overrightarrow{A}\cdot d\overrightarrow{s}\biggr),
\end{equation}
whose regime of validity is discussed in detail in Ref. \citealp{Bernevig-MagField2}.
In this equation, $t_{i,j}$ are hopping amplitudes in the \emph{absence}
of magnetic fields. It is interesting to note that, recently, extensions
of this approximation have been investigated for the case in which
the Wannier functions are delocalized \citep{Bernevig-MagField2}.
Here, we will assume that the Peierls substitution holds. Note that,
in the absence of magnetic fields, there is a rigorous proof that,
for one-dimensional systems, the Wannier functions decay exponentially
\citep{Wannier1D}, as is also the case for two- and three-dimensional
trivial band insulators \citep{Wannier2D3D}.

Let us assume that the uniform magnetic field, $\vec{B}=\nabla\times\vec{A},$
is oriented along the $z$-direction. By gauge freedom, we can choose
the Landau gauge, i.e., $\overrightarrow{A}=-By\hat{x}$. As an illustration,
let us explicitly show how to calculate the hoppings for the case
$D=1$. We have $\tilde{t}_{h,1}^{+}(\vec{B})=t_{h}e^{\frac{iq}{\hbar}\varint_{C_{1}}\overrightarrow{A}\cdot d\overrightarrow{s_{1}}}$,
$\tilde{t}_{h,1}^{-}(\vec{B})=t_{h}e^{-\frac{iq}{\hbar}\varint_{C_{3}}\overrightarrow{A}\cdot d\overrightarrow{s_{3}}}$,
$\tilde{t}_{v,1}(\vec{B})=t_{v}e^{\frac{iq}{\hbar}\varint_{C_{2}}\overrightarrow{A}\cdot d\overrightarrow{s_{2}}}$,
$\tilde{t}_{d,1}^{+}(\vec{B})=t_{d}e^{\frac{iq}{\hbar}\varint_{C_{5}}\overrightarrow{A}\cdot d\overrightarrow{s_{5}}}$,
and $\tilde{t}_{d,1}^{-}(\vec{B})=t_{d}e^{\frac{iq}{\hbar}\varint_{C_{6}}\overrightarrow{A}\cdot d\overrightarrow{s_{6}}}$,
where the integration paths are presented in Fig. \ref{fig:2leg}(b).
It is straightforward to show that $\tilde{t}_{h,1}^{\pm}=t_{h}e^{\pm i\theta}$,
$\tilde{t}_{d,1}^{\pm}=t_{d}$ and $\tilde{t}_{v}=t_{v}$, where $\theta\equiv\frac{q}{\hbar}\frac{\Phi}{2}$
and $\Phi=\iint_{S}\overrightarrow{B}\cdot\hat{n}da$ is the magnetic
flux through the plaquette. For $D>1$, the evaluation of the hopping
amplitudes is similar, and it is also easy to show that $\tilde{t}_{h,\ell}^{\pm}=\frac{t_{h}}{\ell^{\nu}}e^{\pm i\theta\ell}$
and $\tilde{t}_{d,\ell}^{\pm}=\frac{t_{d}}{\ell^{\nu}}$.

Therefore, the Hamiltonian of the two-leg free-fermion ladder with
long-range hopping \emph{in the presence of a uniform magnetic field}
is given by 
\begin{gather}
H(\theta)=-\sum_{n=1}^{L}\sum_{\ell=1}^{D}\Biggl[\frac{t_{h}}{\ell^{\nu}}\Bigl(e^{i\theta\ell}a_{n}^{\dagger}a_{n+\ell}+e^{-i\theta\ell}b_{n}^{\dagger}b_{n+\ell}+h.c.\Bigr)\nonumber \\
+\frac{t_{d}}{\ell^{\nu}}\Bigl(a_{n}^{\dagger}b_{n+\ell}+b_{n}^{\dagger}a_{n+\ell}+h.c.\Bigr)\label{eq:HB}\\
+t_{v}\Bigl(a_{n}^{\dagger}b_{n}+h.c.\Bigr)\Biggr]\textrm{.}\nonumber 
\end{gather}
For $D=1,$ the Hamiltonian reduces to the well-known Creutz model
\citep{PRL-CreutzModel}. The above Hamiltonian can be easily diagonalized
by a Fourier transform. For the sake of completeness, we present the
main steps below. First, we introduce the new fermionic operators
$A_{k}$ and $B_{k}$ by 
\begin{equation}
a_{n}^{\dagger}=\frac{1}{\sqrt{L}}\sum_{k}e^{-ikn}A_{k},\mbox{ and }b_{n}^{\dagger}=\frac{1}{\sqrt{L}}\sum_{k}e^{-ikn}B_{k},\label{eq:Fourier}
\end{equation}
where the momenta are $k=k_{n}=\frac{2\pi}{L}n$, $n=0,1,2,\dots,L-1$.
In terms of $A_{k}$ and $B_{k}$, the Hamiltonian is given by

\begin{eqnarray*}
H & = & \sum_{k}\left(\begin{array}{cc}
A_{k}^{\dagger} & B_{k}^{\dagger}\end{array}\right)\left(\begin{array}{cc}
2t_{h}C_{k+\theta} & 2t_{d}C_{k}+t_{v}\\
2t_{d}C_{k}+t_{v} & 2t_{h}C_{k-\theta}
\end{array}\right)\left(\begin{array}{c}
A_{k}^{\phantom{\dagger}}\\
B_{k}^{\phantom{\dagger}}
\end{array}\right),
\end{eqnarray*}
where 
\begin{eqnarray}
C_{k}=C_{k}(\nu,D) & = & \sum_{\ell=1}^{D}\ell^{-\nu}\cos\left(k\ell\right).\label{eq:C}
\end{eqnarray}
Note that the above Hamiltonian has the same form as the one presented
in Sec. \ref{sec:twobandModel}, so we can use all the results derived
there. We only need to identify that $M_{12}(k)=M_{21}(k)=2t_{d}C_{k}+t_{v}$,
$M_{11}(k,\theta)=2t_{h}C_{k+\theta}$, and $M_{22}(k,\theta)=2t_{h}C_{k-\theta}$.
In the next subsection, we use the results of the previous section
to investigate the DQPTs of the two-leg Creutz ladder.

It is worth noting that, for some special values of $\nu$ and $D$,
the functions $C_{k}$ can also be expressed in terms of several well-known
functions. For instance, for $D=1$, $C_{k}(\nu,1)=\cos k$, while
for $\nu=0$ we have 
\begin{equation}
C_{k}(0,D)=\frac{\sin\left[(D+1)k/2\right]}{\sin\Bigl(k/2\Bigr)}\cos(Dk/2)-1.
\end{equation}
For $D=\infty$, $C_{k}(1,\infty)=\Re\Bigl\{-\ln(1-e^{ik})\Bigr\}$,
for $\nu=1$, whereas for arbitrary values of $\nu$, $C_{k}(\nu,\infty)=\Re\Bigl\{ Li_{\nu}\bigl(e^{ik}\bigr)\Bigr\}$,
where $\text{Li}_{\nu}(z)$ is the polylogarithm function of order
$\nu$.

\begin{figure}
\begin{centering}
\includegraphics[scale=0.55]{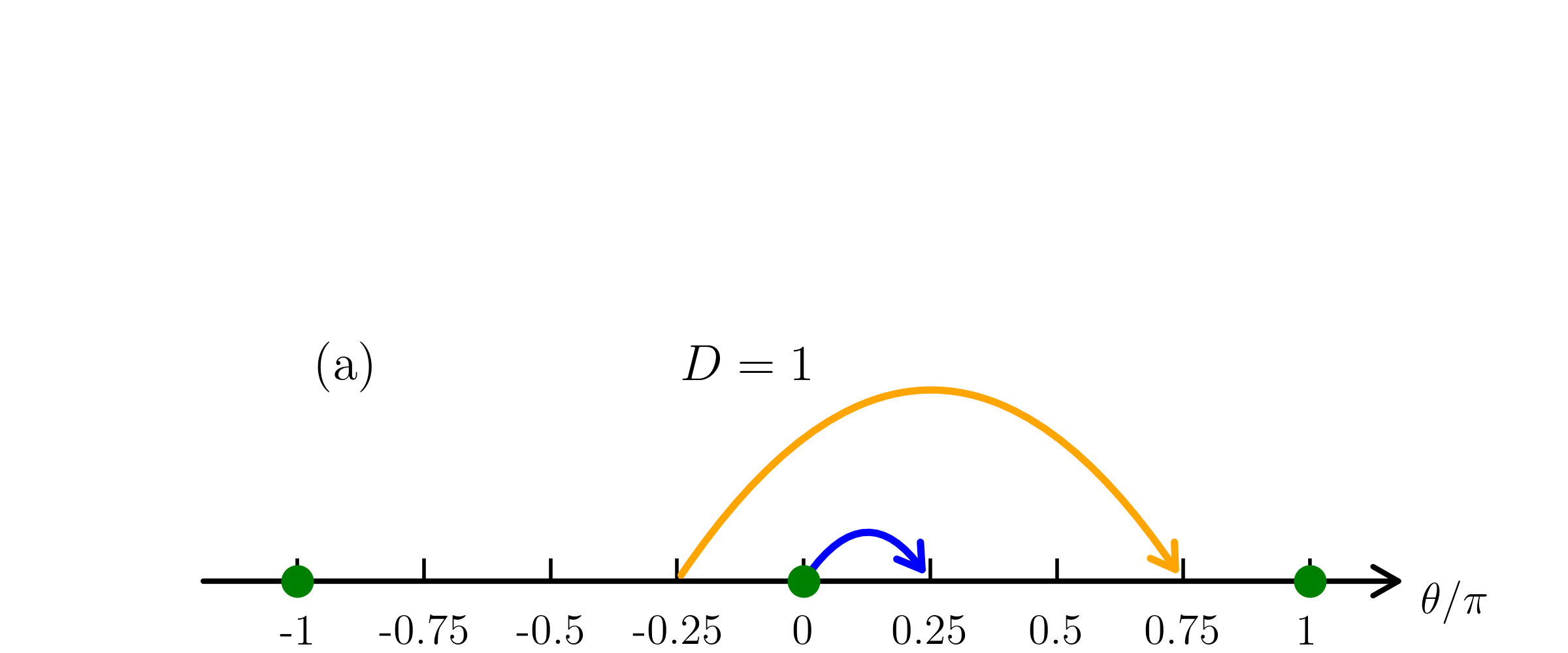}
\par\end{centering}
\vspace*{0.5cm}

\begin{centering}
\includegraphics[scale=0.55]{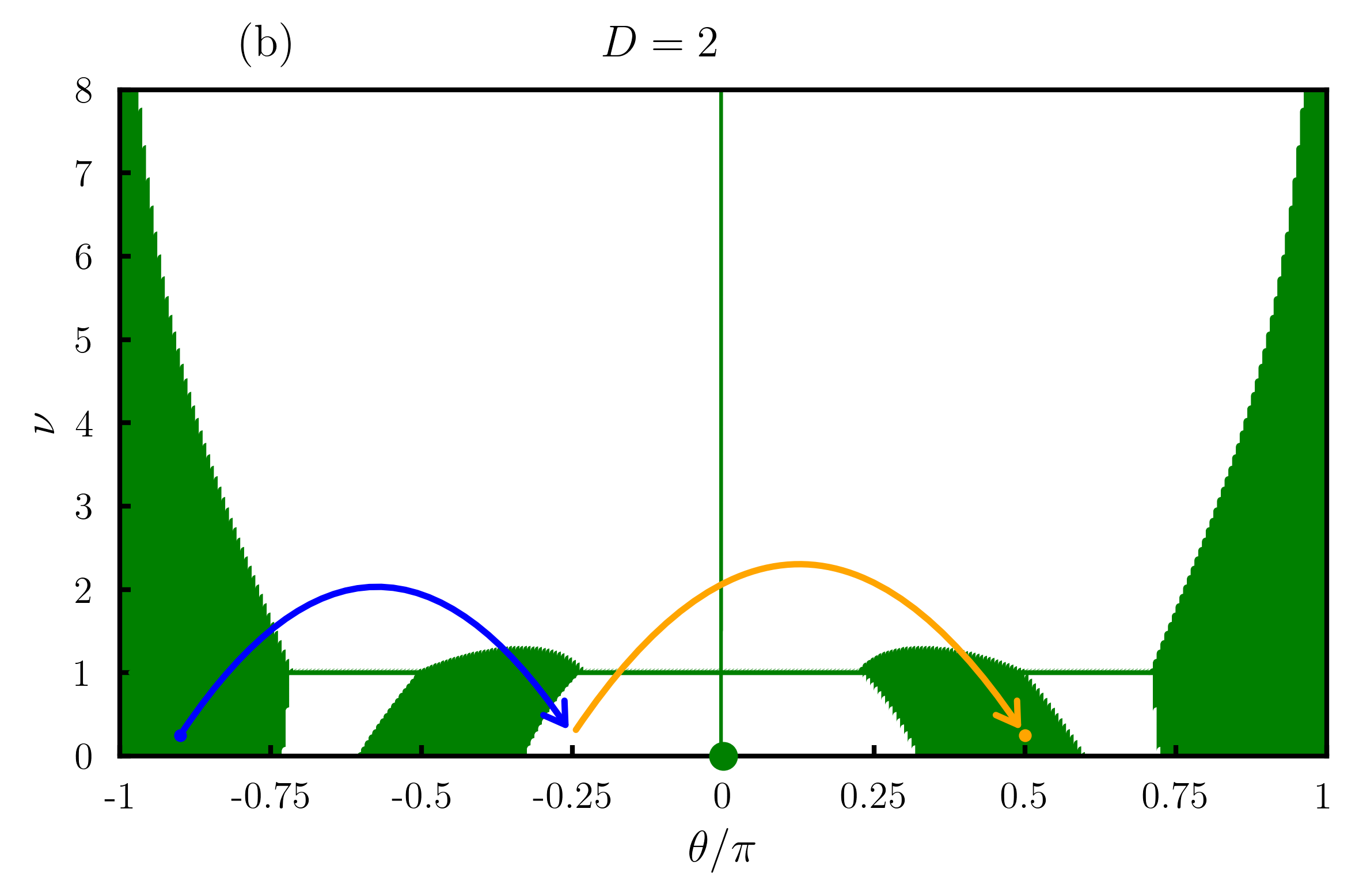}
\par\end{centering}
\begin{centering}
\includegraphics[scale=0.55]{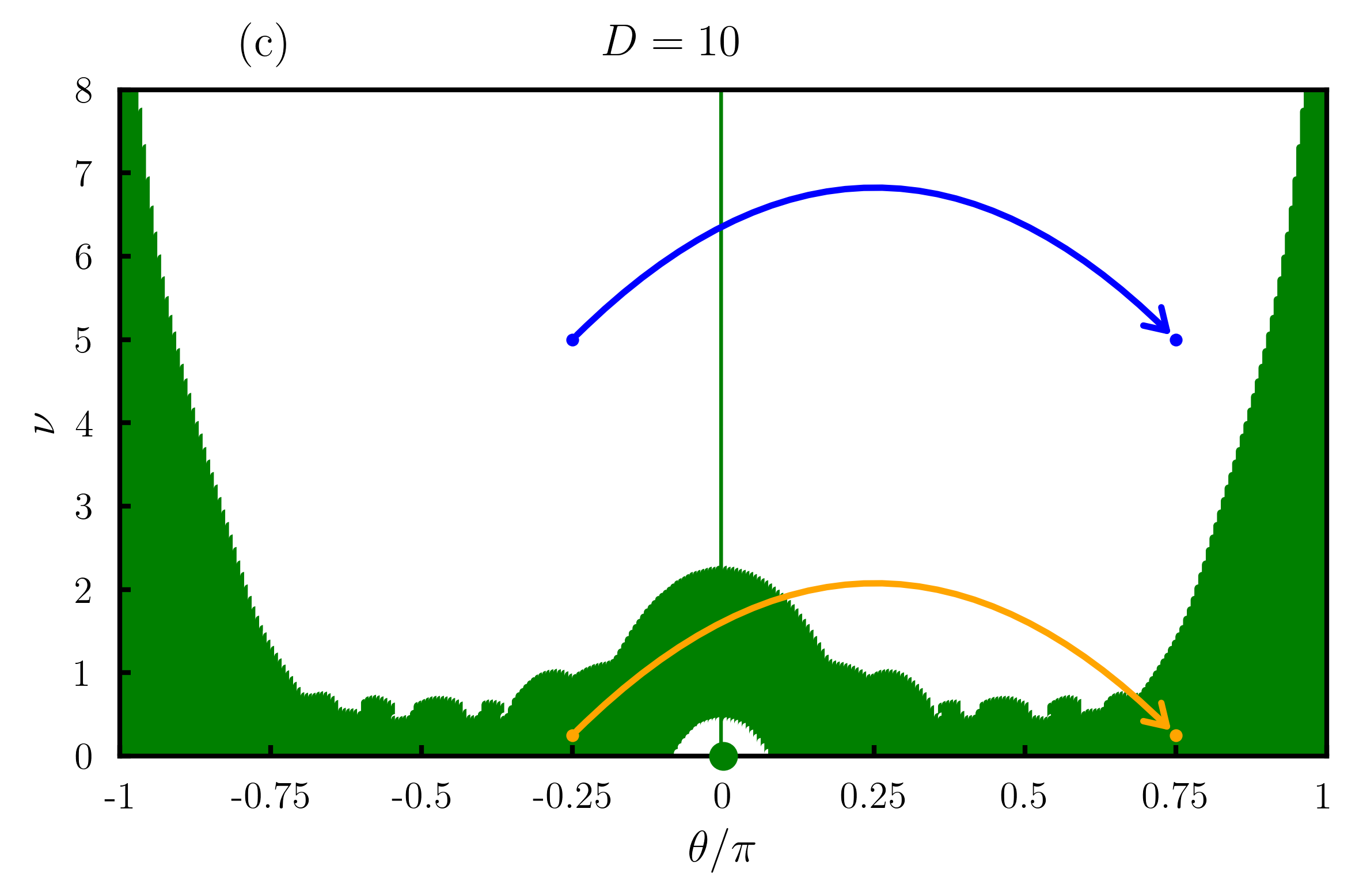}
\par\end{centering}
\caption{Critical regions for $D=1$, $D=2$, and $D=10$. In the green region,
the gap is zero (critical region), while the white region is gapped.
The arrows indicate the quench parameters considered in Figs. \ref{fig:D1}
and \ref{fig:freeEnerDs}. \label{fig:gapsD}}
\end{figure}

One of the key ingredients for understanding the DQPTs in the Creutz
model is the dispersion {[}see Eq. (\ref{eq:dispertion2band}){]}
given by 
\begin{eqnarray}
\omega_{k}^{\pm}(\theta) & = & -t_{h}(C_{k+\theta}+C_{k-\theta})\nonumber \\
 &  & \pm\sqrt{t_{h}^{2}(C_{k+\theta}-C_{k-\theta})^{2}+(2t_{d}C_{k}+t_{v})^{2}}.\label{eq:dispenstionCreutz}
\end{eqnarray}
The numerical results presented in the following refer to the Creutz
model at half-filling, i.e., at a density $\rho=1/2$, with $t_{v}=t_{h}=t_{d}=1$.
Note that, from Eq. (\ref{eq:dispenstionCreutz}), it is easy to show
that if $\theta\ne0,\pm\pi$ $(\mod2\pi)$, the system is always gapped
in equilibrium (see discussion in the following). On the other hand,
the gap is zero if there exists a momentum $k_{c}^{eq}$ and a phase
$\theta_{c}^{eq}$ that simultaneously satisfy the equations

\begin{eqnarray}
C_{k_{c}^{eq}}(\nu,D) & = & -\frac{t_{v}}{2t_{d}},\label{eq:thetac1}\\
C_{k_{c}^{eq}+\theta_{c}^{eq}} & = & C_{k_{c}^{eq}-\theta_{c}^{eq}}.\label{eq:thetac2}
\end{eqnarray}
Using the fact that $C_{k}=C_{k+2\pi}$ and $C_{k}=C_{-k}$, we find
that Eq. (\ref{eq:thetac2}) is satisfied for $\theta_{c}^{eq}=0,\pm\pi$.
We then need to find values of $k_{c}^{eq}(D,\nu)$ by solving Eq.
(\ref{eq:thetac1}). For the case $D=1$, if $t_{v}/t_{d}\le2$, the
solutions are $k_{c,\pm}^{eq}(1,\nu)=\pi\pm\arccos(\frac{t_{v}}{2t_{d}})$
\citep{PRL-CreutzModel,dynamicDelgadoPhysRevB.99.054302}. In Fig.
\ref{fig:gapsD}(a), we present these two critical values of $\theta_{c}^{eq}$
in the parameter space. 

For $D=2$, if $t_{v}/t_{d}\le2^{\nu-2}+2^{1-3\nu}$, the momenta
that satisfy Eq. (\ref{eq:thetac1}),\footnote{We may also have a vanishing gap, for arbitrary values of $\theta_{c}^{eq}$,
for some very specific values of the hopping amplitudes. For instance,
for the couplings $t_{v}/t_{d}=-1/2\sum_{\ell}^{D}1/\ell^{\nu}$ at
$k_{c}^{eq}=0$ and $t_{v}/t_{d}=-1/2\sum_{\ell}^{D}(-1)^{\ell}/\ell^{\nu}$
at $k_{c}^{eq}=\pm\pi$.} are given by 
\begin{equation}
k_{c,\pm}^{eq}(2,\nu)=\arccos\left[-2^{\nu-2}\pm2^{\nu-2}\sqrt{1+2^{3-2\nu}-2^{2-\nu}t_{v}/t_{d}}\right],
\end{equation}
where the solution $k_{c,-}^{eq}(2,\nu)$ exists only if there are
values of $\nu$ satisfying $\cos^{2}\left(k_{c,-}^{eq}(2,\nu)\right)\le1$.
We now need to find the values of $\theta_{c}^{eq}$ that satisfy
Eq. (\ref{eq:thetac2}). For $\theta_{c}^{eq}=0,\pm\pi$, there is
always a solution to Eq. (\ref{eq:thetac2}). Besides these critical
values, we verify numerically that several other values of $\theta_{c}^{eq}$
also satisfy Eq. (\ref{eq:thetac2}). 

From now on, we set $t_{v}=t_{h}=t_{d}=1$. In Fig. \ref{fig:gapsD}(b),
we present the critical values of $\theta_{c}^{eq}$ in the parameter
space for $D=2$. Note that there are forbidden energy regions, depending
on the value of the magnetic flux $\theta$, which resemble the Hofstadter
butterfly \citep{HofstadterButterfly}. The critical region for $D=10$
is also presented in Fig. \ref{fig:gapsD}(c). Finally, for $\theta=0,\pm\pi$,
we have numerically verified that, for several values of $D$ and
$\nu$, there exists at least one value of $k_{c}^{eq}$ satisfying
Eq. (\ref{eq:thetac2}). Based on this observation and the discussion
above, we conjecture that, for $\theta_{c}^{eq}=0,\pm\pi$, the Creutz
model with long-range hopping has a vanishing gap when $t_{v}=t_{h}=t_{d}=1$.

\subsection{DQPT Results of the Creutz Model \label{subsec:Results} }

As mentioned before, the two key quantities needed to investigate
the DQPTs are the dispersion given in Eq. (\ref{eq:dispenstionCreutz})
and the phase $\gamma_{k,\theta}$ {[}see Eq. (\ref{eq:phasegamma}){]},
obtained from

\begin{equation}
\tan(\gamma_{k,\theta})=-\frac{2t_{d}C_{k}+t_{v}}{t_{h}\bigl(C_{k+\theta}-C_{k-\theta}\bigr)}.\label{eq:phasegammaCreutz}
\end{equation}
All relevant quantities, such as the dynamical free energy and the
YLF zeros of the Creutz model, are expressed in terms of the dispersion
$\omega_{k}^{\pm}(\theta)$ given by (\ref{eq:dispenstionCreutz})
and the phase $\gamma_{k,\theta}$ obtained from Eq. (\ref{eq:phasegammaCreutz}).
We use these results to evaluate the dynamical free energy $f(\texttt{t})$
and the critical times $\texttt{t}_{c}$ of the Creutz model. In what
follows, we consider quenches from $\theta$ to $\tilde{\theta}$,
keeping $\nu$ and $D$ fixed.

Before proceeding, let us first determine the region of coupling-parameter
space where DQPTs may occur in the two-leg Creutz model. DQPTs occur
only if there exists a $k^{\star}$ that satisfies {[}see Eq. (\ref{eq:YLZ-tau0}){]}

\begin{equation}
\left(2t_{d}C_{k^{\star}}+t_{v}\right)^{2}=-t_{h}^{2}\left(C_{k^{\star}+\theta}-C_{k^{\star}-\theta}\right)\left(C_{k^{\star}+\tilde{\theta}}-C_{k^{\star}-\tilde{\theta}}\right).\label{eq:kStarCreutz}
\end{equation}
We only need to find values of $k^{\star}$ in the interval $[0,\pi]$,
since all relevant quantities {[}see, for instance, Eqs. (\ref{eq:tc}),
(\ref{eq:dispenstionCreutz}), and (\ref{eq:phasegammaCreutz}){]}
depend on the momentum $k$ through $C_{k}$, which is symmetric about
$k=\pi.$ Note also that the left-hand side of the above equation
is positive; thus DQPTs occur \emph{only if} $\left(C_{k^{\star}+\theta}-C_{k^{*}-\theta}\right)\left(C_{k^{\star}+\tilde{\theta}}-C_{k^{\star}-\tilde{\theta}}\right)\le0$.

\subsubsection*{Case $D=1$}

\begin{figure}
\begin{centering}
\includegraphics[scale=0.55]{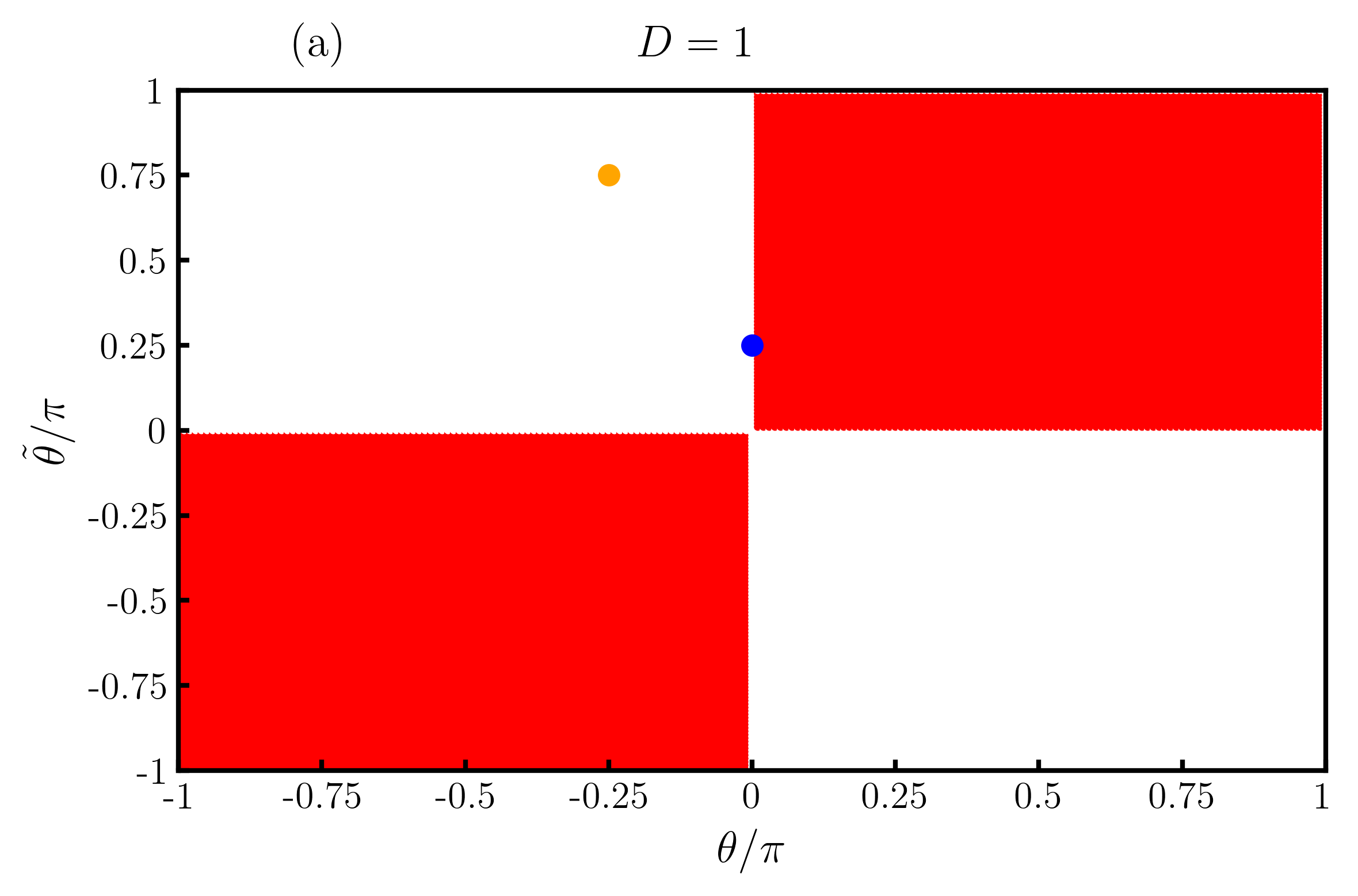}
\par\end{centering}
\vspace*{0.5cm}

\begin{centering}
\includegraphics[scale=0.55]{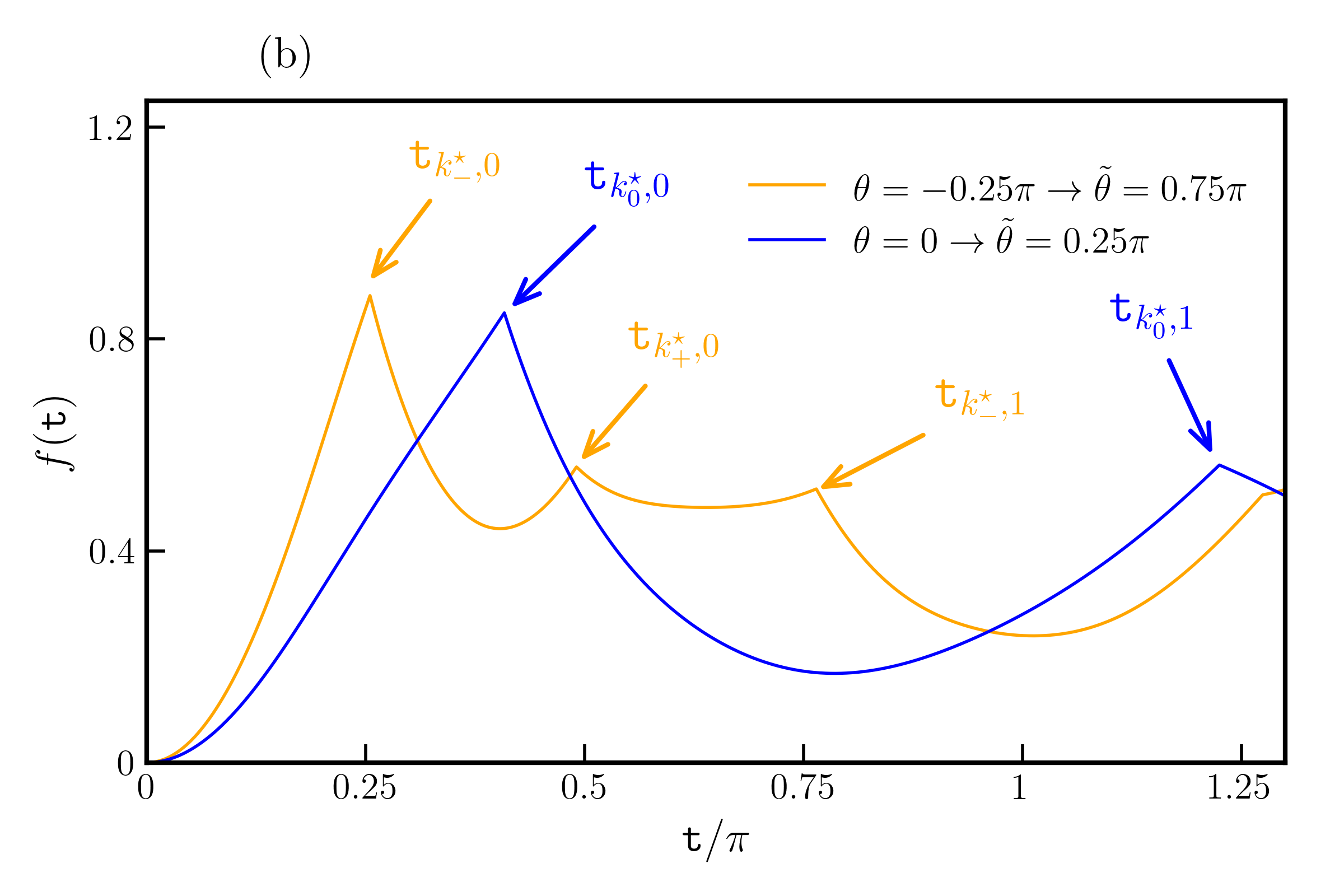}
\par\end{centering}
\caption{Results for $D=1$. (a) Diagram showing the parameter region where
DQPTs occur (white region) for a quench from $\theta$ to $\tilde{\theta}$.
In the red region, DQPTs do not occur. The dots in this diagram correspond
to the two quench parameters used in (b). (b) The dynamical free energy
$f(\texttt{t})$ vs. $\texttt{t}/\pi$ for a system of size $L=40000$
and two quenches (see legend). The arrows indicate the cusp positions
at the critical times $t_{k_{\pm}^{\star},m}$ and $t_{k_{0}^{\star},m}$
(see text). The momenta $k_{\pm}^{\star}$ and $k_{0}^{\star}$ are
given by Eqs. (\ref{eq:kstarD1a}) and (\ref{eq:kstarD1b}), respectively.
\label{fig:D1}}
\end{figure}

For $D=1$, we have 
\begin{equation}
\left(C_{k^{\star}+\theta}-C_{k^{\star}-\theta}\right)\left(C_{k^{\star}+\tilde{\theta}}-C_{k^{\star}-\tilde{\theta}}\right)=4\sin^{2}k^{\star}\sin\theta\sin\tilde{\theta}
\end{equation}
which implies that DQPTs occur if $\theta$ and $\tilde{\theta}$
have opposite signs. For clarity and comparison with the results for
different values of $D$, we present in Fig. \ref{fig:D1}(a) the
region where DQPTs occur for $D=1$. Note that DQPTs in the case $D=1$
occur when the \emph{quench crosses} the equilibrium critical point
at $\theta_{c}^{eq}=0$, as previously reported in Ref. \onlinecite{dynamicDelgadoPhysRevB.99.054302}.
In this case, it is easy to show that $k^{\star}$ is given by 
\begin{equation}
k_{\pm}^{\star}=\arccos\left[\frac{-1\pm\sqrt{\sin\theta\sin\tilde{\theta}\left(4\sin\theta\sin\tilde{\theta}-3\right)}}{2(\sin\theta\sin\tilde{\theta}-1)}\right],\label{eq:kstarD1a}
\end{equation}
Consequently, we have two families of critical times, given by $\texttt{t}_{k_{\pm}^{\star},m}=\frac{2\pi}{\Delta\tilde{E}_{k_{\pm}^{\star}}}\Bigl(m+\frac{1}{2}\Bigr)$.
The dynamical free energy is non-analytic at these critical times
(see Fig. \ref{fig:D1}).

It is worth mentioning that DQPTs can also occur when the quench starts
from the equilibrium critical point, i.e., when $\theta=0$ and $\tilde{\theta}\ne0$.
In this case, we have only one family of critical times $\texttt{t}_{k_{0}^{\star},m}$,
where

\begin{equation}
k_{0}^{\star}=2\pi/3.\label{eq:kstarD1b}
\end{equation}
This family of critical times was not reported in Ref. \onlinecite{dynamicDelgadoPhysRevB.99.054302}.
Note that for a quench from $\theta\ne0$ to $\tilde{\theta}=0$,
$\texttt{t}_{k_{0}^{\star},m}$ diverges, since $\Delta\tilde{E}_{k_{0}^{\star}}=0$.

Although DQPTs in the Creutz model with $D=1$ were previously investigated
in Ref. \onlinecite{dynamicDelgadoPhysRevB.99.054302}, we present,
for completeness and for the sake of comparison, some results in Fig.
\ref{fig:D1}. In Fig. \ref{fig:D1}(a), we show the parameter region
in which the DQPTs occur, while in Fig. \ref{fig:D1}(b), we report
representative examples of the dynamical free energy $f(\texttt{t})$,
considering two quenches: one from $\theta=0$ to $\tilde{\theta}=0.25\pi$,
and another from $\theta=-0.25\pi$ to $\tilde{\theta}=0.75\pi$. 

\begin{figure}
\vspace*{0.25cm}

\begin{centering}
\includegraphics[scale=0.55]{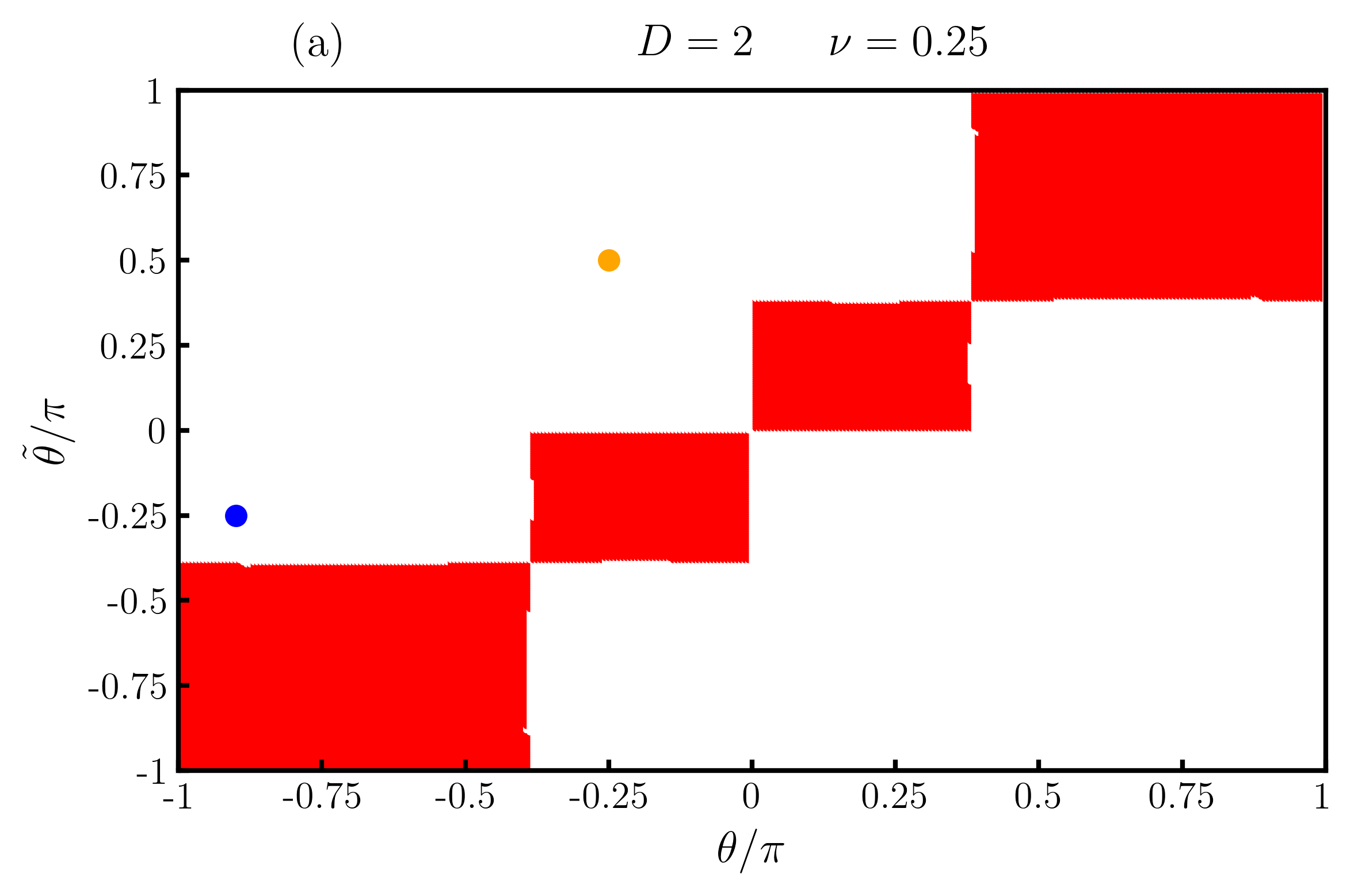}
\par\end{centering}
\vspace*{0.25cm}

\begin{centering}
\includegraphics[scale=0.55]{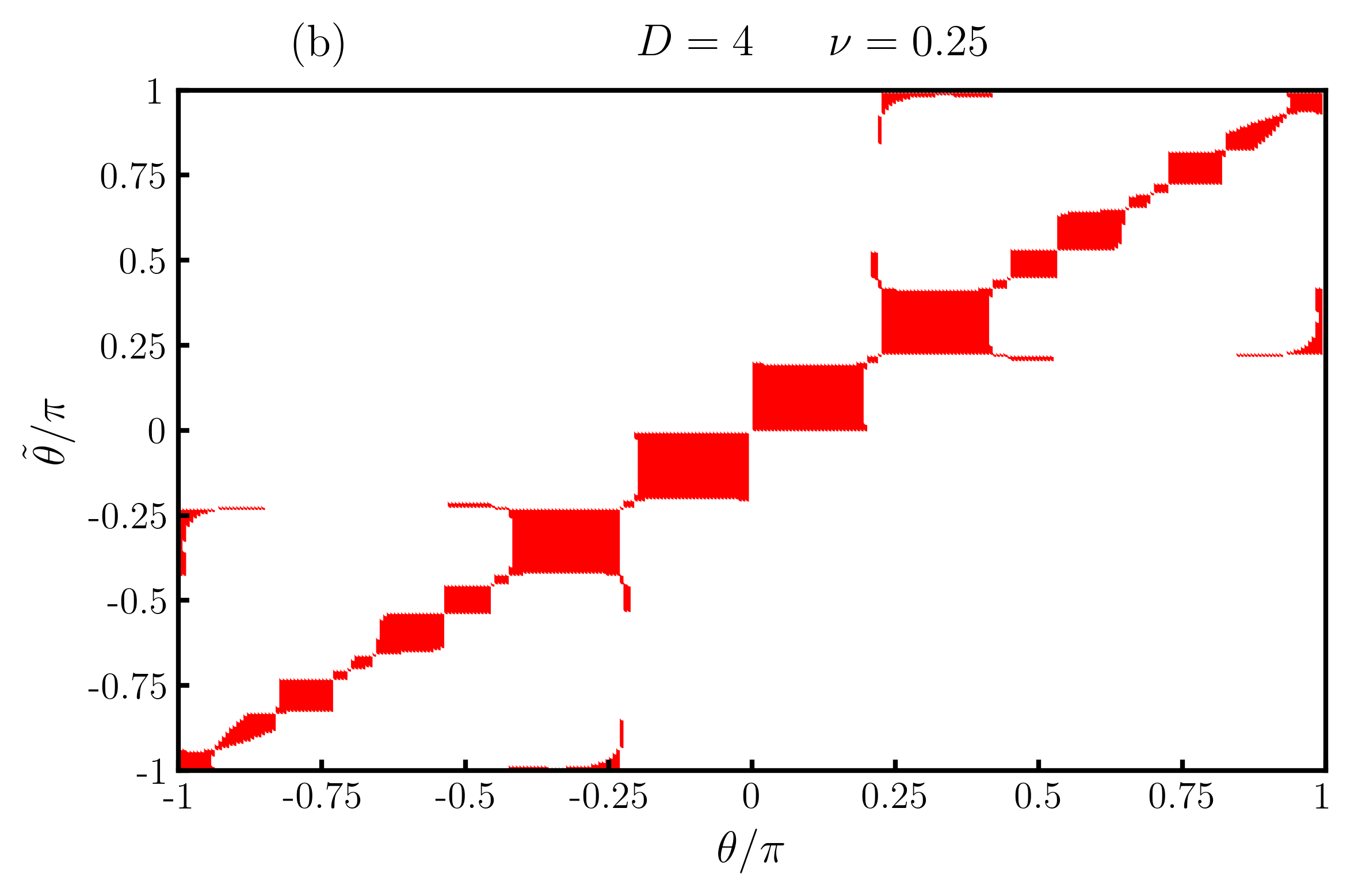}
\par\end{centering}
\vspace*{0.25cm}

\begin{centering}
\includegraphics[scale=0.55]{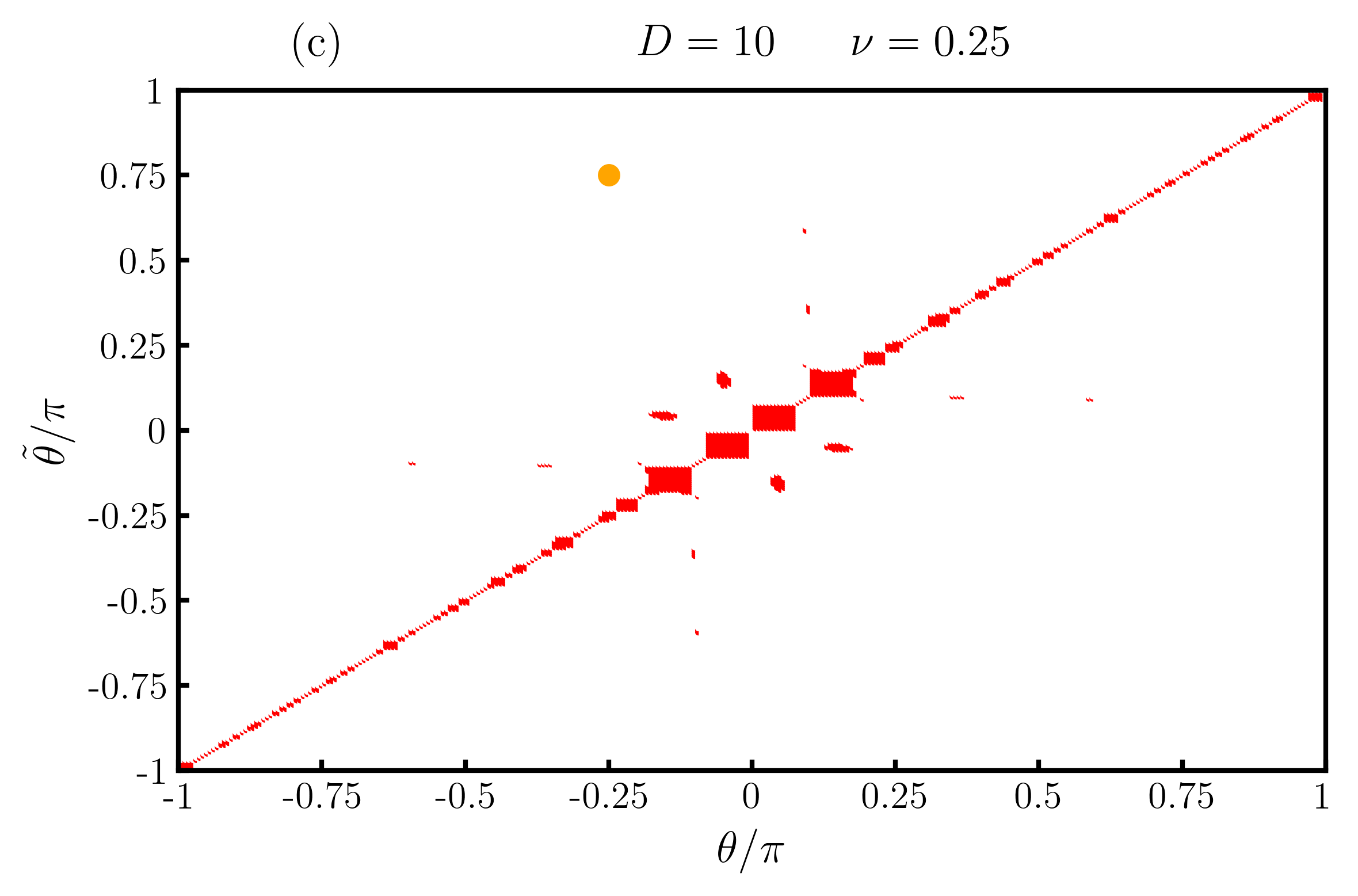}
\par\end{centering}
\caption{Parameter region where DQPTs occur (white region) for $\nu=0.25$;
in the red region, DQPTs do not occur. Results for $D=2$, $D=4$,
and $D=10$ are presented in (a), (b), and (c), respectively. The
dots in (a) and (c) indicate two of the quench parameters considered
in Fig. \ref{fig:freeEnerDs} for $\nu=0.25$. \label{fig:Ds}}
\end{figure}

\subsubsection*{Case $D>1$}

For $D>1$, DQPTs may occur for \emph{quenches that do not cross}
the equilibrium critical point $\text{\ensuremath{\theta_{c}^{eq}}}$,
depending on the exponent $\nu,$ as we discuss below. First, note
that for $\nu\rightarrow\infty$, $C_{k}\rightarrow\cos k$. Therefore,
for any $D$ and $\nu\gg1$, we obtain the same results as in the
case of $D=1$. In this case, DQPTs occur when the \emph{quench crosses}
the equilibrium critical point at $\theta_{c}^{eq}=0$. On the other
hand, when $\nu\lesssim1$, the scenario is quite different, particularly
for large values of $D$. To illustrate this, we solve numerically
Eq. (\ref{eq:kStarCreutz}) to obtain the parameter region where DQPTs
occur for quenches from $\theta$ to $\tilde{\theta}$, keeping $\nu$
and $D$ fixed. In Fig. \ref{fig:Ds}, we show this region for $\nu=0.25$
and some values of $D$. Interestingly, for $D>1$, DQPTs may occur
even when the quench traverses a gapped region without crossing an
equilibrium critical point {[}see, for example, Figs. \ref{fig:Ds}(a)
and \ref{fig:gapsD}(b) for the case $D=2${]}. Moreover, note that
as we increase $D$, DQPTs appear to occur almost everywhere, and
they can occur \emph{without the quench crossing} an equilibrium critical
point. We observe that, for certain parameters, the DQPT region does
not necessarily coincide with the equilibrium phase transition region.
It is worth mentioning that there are several other examples in the
literature where DQPTs occur without crossing an equilibrium critical
point \citep{DynamSirkerPRB2014,DynamicVajnaPRB2014,FirtOderDynanimcPRL,PRBVajnaDynamic,DynamicLongRangePRB2017,PRLHeylLongRangDynamic,SciRepJafari,NetoRafaelXavierPRBL2022}.

\begin{figure}
\begin{centering}
\includegraphics[scale=0.55]{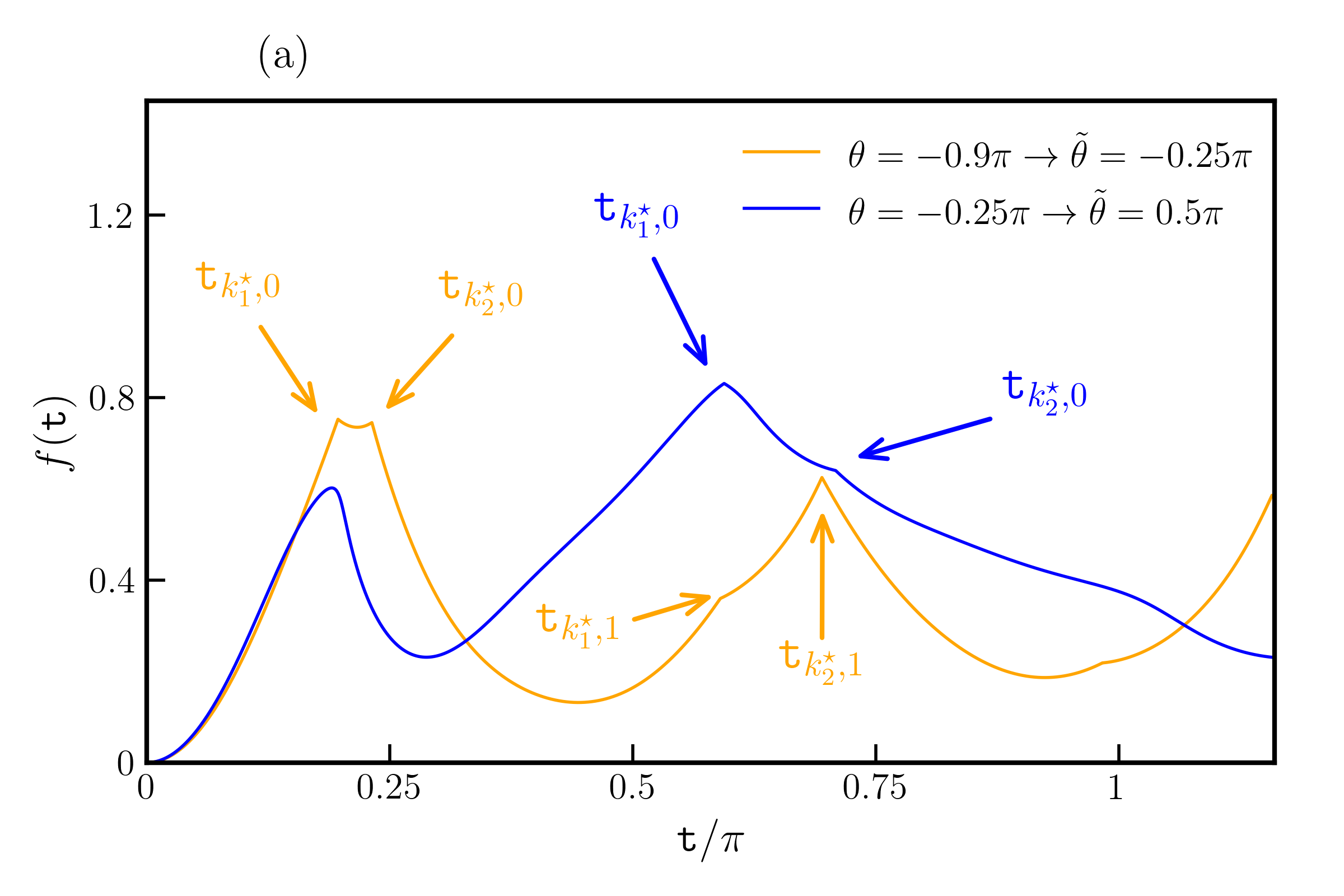}
\par\end{centering}
\vspace*{0.5cm}

\begin{centering}
\includegraphics[scale=0.55]{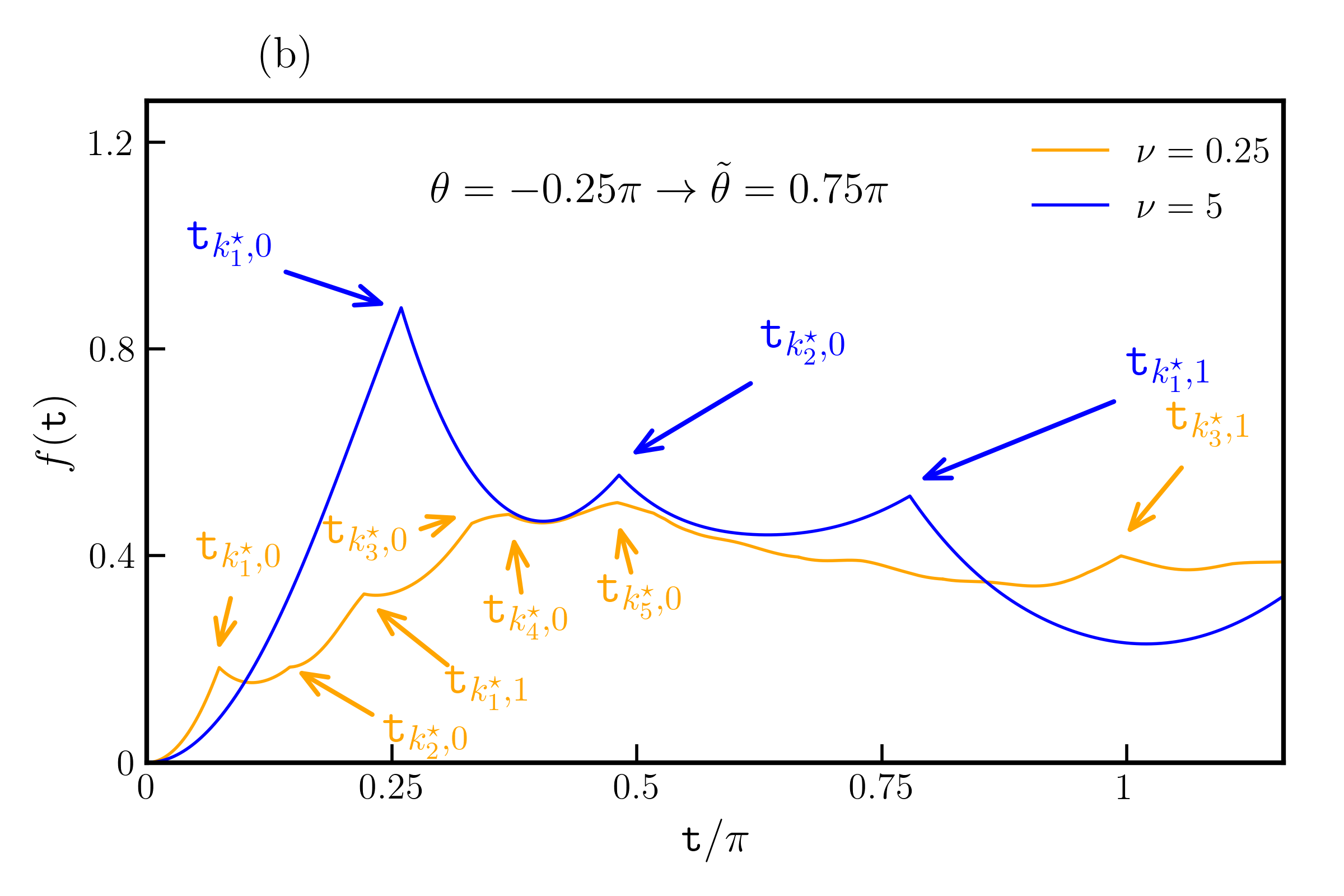}
\par\end{centering}
\vspace*{0.5cm}

\begin{centering}
\includegraphics[scale=0.55]{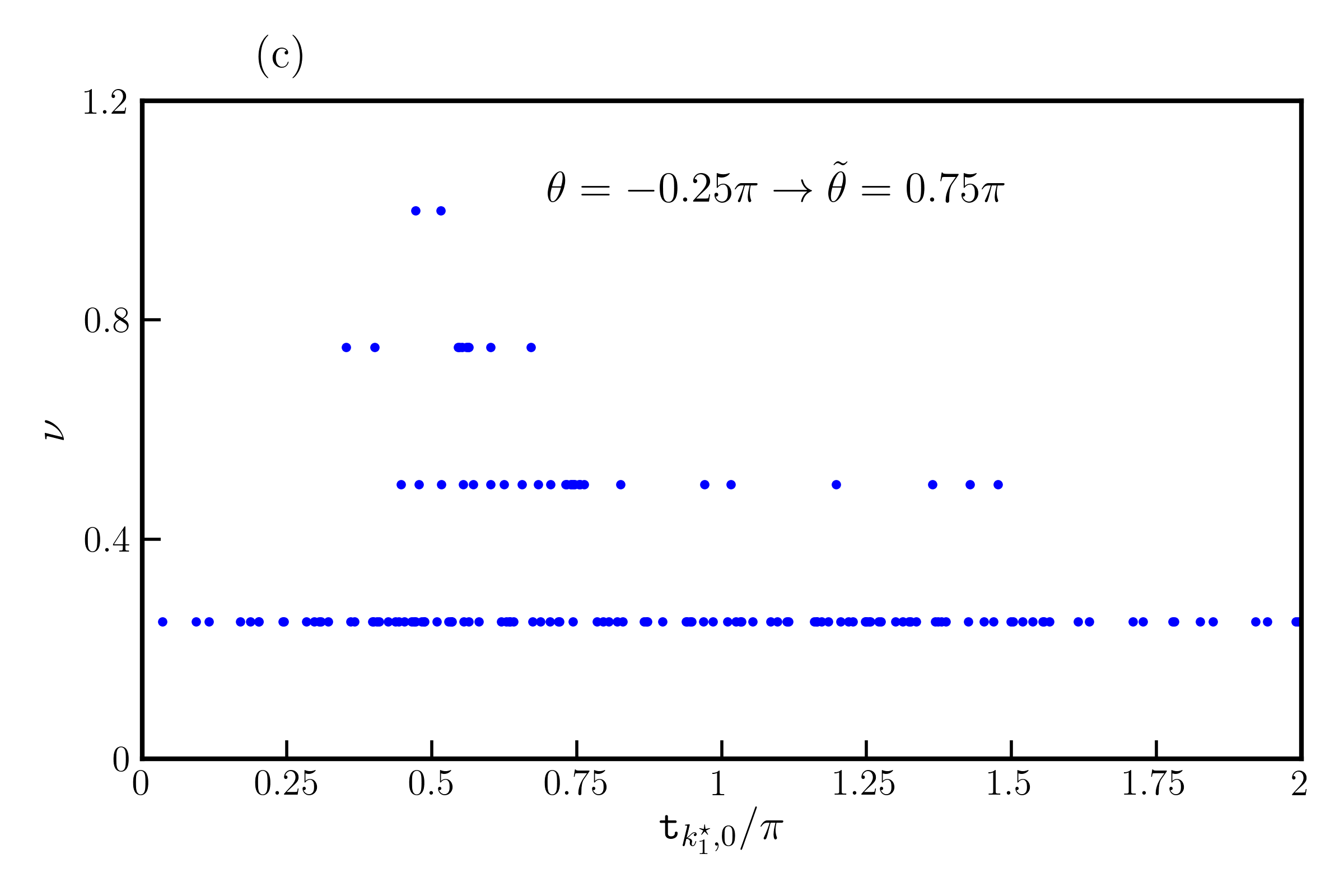}
\par\end{centering}
\caption{(a) The dynamical free energy $f(\texttt{t})$ vs. $\texttt{t}/\pi$
for a system of size $L=40000$ with $D=2$ and some quenches (see
legend). The arrows indicate the cusp positions at the critical times
$\texttt{t}_{k_{i}^{\star},m}$ (see text). (b) Same as in (a), but
for $D=10$ and a quench from $\theta=-0.25\pi$ to $\tilde{\theta}=0.75\pi$
and two values of $\nu$ (see legend). (c) The earliest critical time
$\texttt{t}_{k_{1}^{\star},0}$ for some values of $\nu$ for a quench
from $\theta=-0.25\pi$ to $\tilde{\theta}=-0.75\pi$ with $D=100$.
\label{fig:freeEnerDs}}
\end{figure}

It is important to mention that\emph{ not all times at which} $f(\texttt{t})$\emph{
reaches a local maximum} correspond to non-analyticities at those
times, as we illustrate below. In Fig. \ref{fig:freeEnerDs}(a), we
present the dynamical free energy for two quenches with $D=2$: one
from $\theta=-0.9\pi$ to $\tilde{\theta}=-0.25\pi$ (orange curve)
and another from $\theta=-0.25\pi$ to $\tilde{\theta}=0.5\pi$ (blue
curve). In both cases, we have two wave vectors that solve Eq. (\ref{eq:kStarCreutz}).
These wave vectors give rise to two families of critical times, $\texttt{t}_{k_{i}^{\star},m}=\frac{2\pi}{\Delta\tilde{E}_{k_{i}^{\star}}}\Bigl(m+\frac{1}{2}\Bigr)$,
$i=1,2$. The arrows in this figure indicate the positions of the
critical times $\texttt{t}_{k_{i}^{\star},m}$. Note that, for the
orange curve, all local maxima correspond to critical times. On the
other hand, for the blue curve, the first local maximum is not associated
with any of the critical times $\texttt{t}_{k_{i}^{\star},m}$, i.e.,
$f(\texttt{t})$ is analytic at this local maximum. Similar mismatches
between extrema of the Loschmidt rate and critical times have recently
been reported in long-range extensions of the SSH model \citep{Sacramento2}.

In general, the non-analyticities of $f(\texttt{t})$ emerge only
in infinite systems (although for some very particular cases they
may appear in finite systems; see, for instance, Refs. \onlinecite{DynamSirkerPRB2014,KarraschPhysRevB.87.195104,DQFTfinitesystem,ZHU2024129455,Xavier-DQPT-LR}).
Thus, the examples discussed above demonstrate that determining the
non-analyticities of $f(\texttt{t})$ from experimental data can be
quite challenging, because experiments typically access $f(\texttt{t})$
only for relatively small system sizes. Due to this limitation, smeared
local maxima may appear, which are not necessarily associated with
true non-analyticities in $f(\texttt{t})$. Therefore, theoretical
results are highly desirable to corroborate experimental evidence
of non-analyticities in $f(\texttt{t})$.

For large values of $\nu$ and any value of $D$, we expect that the
dynamical free energy $f(\texttt{t})$ is similar to that for $D=1$,
as mentioned before. Indeed, this is the case, as illustrated in Fig.
\ref{fig:freeEnerDs}(b) for $D=10$ and $\nu=5$, for a quench from
$\theta=-0.25\pi$ to $\tilde{\theta}=0.75\pi$. In this case, we
again have only two wave vectors, $k_{i}^{\star}$ ($i=1,2)$, that
solve Eq. (\ref{eq:kStarCreutz}). Note the similarity between this
figure and the one presented in Fig. \ref{fig:D1}(b) for the same
quench.

Another important observation is that, as we decrease $\nu$, the
number of solutions to Eq. (\ref{eq:kStarCreutz}) increases. In particular,
if we consider the same quench used in Fig. \ref{fig:freeEnerDs}(b),
but with $\nu=0.25$, we obtain 12 values of $k_{i}^{\star}$ that
solve Eq. (\ref{eq:kStarCreutz}). Consequently, for large values
of $D$ and small values of $\nu$, the critical times can become
increasingly dense, as illustrated in Fig. \ref{fig:freeEnerDs}(c).
Note that, in this figure, we only show the earliest critical time,
$\texttt{t}_{k_{1}^{\star},0}$. Actually, we have several other critical
times $\texttt{t}_{k_{1}^{\star},m}$, with $m=1,2,3,...$ {[}see
Eq. (\ref{eq:tc}){]}, which lead to a much denser interval of critical
times (not shown in the figure). A similar observation was also made
for the Su-Schrieffer-Heeger (SSH) chain with long-range hopping terms
\citep{Xavier-DQPT-LR}. This suggests that such behavior may be a
general characteristic of models with long-range hopping.

\section{Further discussions and conclusions\label{sec:CONCLUSION}}

We have investigated dynamical quantum phase transitions in the two-leg
Creutz ladder with long-range hopping. We first derived the exact
solution of a generic two-band free-fermion model in momentum space
and obtained closed-form expressions for the Loschmidt amplitude,
the dynamical free energy, and the Yang-Lee-Fisher zeros. This formulation,
expressed directly in terms of eigenstates, energy eigenvalues, and
their overlaps, provides a convenient framework for determining the
critical times associated with DQPTs.

We then applied this framework to the Creutz model with long-range
hopping. We found that the number of solutions to the equation determining
the critical momenta increases as the decay exponent $\nu$ decreases.
For large hopping ranges $D$ and small $\nu$, the corresponding
critical times can become increasingly dense, leading, in the appropriate
regime, to non-analyticities at an increasingly dense set of times.
This behavior is similar to that previously reported for the SSH model
with long-range hopping \citep{Xavier-DQPT-LR} and suggests that
the emergence of densely distributed critical times may be a general
feature of long-range hopping systems.

Our results also show that, for $D>1$, DQPTs need not be associated
with a quench crossing an equilibrium critical point. Long-range hopping
can therefore qualitatively modify the relation between equilibrium
and nonequilibrium critical behavior. Since the Creutz model has been
realized experimentally, these results may also be relevant for future
experimental investigations of DQPTs and the effects of long-range
hopping.
\begin{acknowledgments}
The authors thank U. R. Fischer for bringing Ref. \citep{PhysRevLett.97.200601}
to their attention. J.C.X. acknowledges support from the INCT project
Advanced Quantum Materials, involving the Brazilian agencies CNPq
(Proc. 408766/2024-7), FAPESP (Proc. 2025/27091-3), and CAPES. J.A.S.
thanks CAPES for support. 
\end{acknowledgments}

\bibliographystyle{apsrev4-1}
\phantomsection\addcontentsline{toc}{section}{\refname}\bibliography{/home/jcxavier/FILES/textos/refs_rev4}

\end{document}